
\input amstex
\magnification=\magstep1
\documentstyle{amsppt}
\NoBlackBoxes
\font\BIG=cmbx12 scaled\magstep1

\define\vs{\vskip.2cm}
\define\ve{\varepsilon}

\define\ga{\gamma}
\define\ot{\otimes}

\define\f|{\left |}
\define\r|{\right |}

\define\<{\left <}
\define\>{\right >}
\define\Z{\Bbb Z}
\define\upa{\uparrow}
\define\si{\sigma}
\define\doa{\downarrow}
\define\al{\alpha}
\define\ch{\check}
\define\Del{\Delta}
\pageno=1
\document
\baselineskip=6mm plus 2pt
\vskip.2cm
\TagsOnRight
\parindent 8 mm

\rightline{YITP/K-1114}
\rightline{
cond-mat/9506119
}
\vskip.2in
\centerline{\BIG ``Integrable electron model with correlated hopping}
\centerline{\BIG and quantum supersymmetry'' }
\vskip 1cm
\centerline {\bf by}
\vskip 1cm
\centerline {\bf Mark D. Gould, Katrina E. Hibberd, Jon R. Links}
\centerline {Department of Mathematics, University of Queensland,
Queensland, 4072, Australia}
\centerline {email jrl\@maths.uq.oz.au}
\+\cr  \centerline {and}
\+\cr
\centerline {\bf Yao-Zhong  Zhang}
\centerline {Yukawa Institute for Theoretical Physics, Kyoto University,
Kyoto 606, Japan}
\centerline {email yzzhang\@yukawa.kyoto-u.ac.jp}
\+\cr
\centerline {June, 1995}
\+\cr
\+\cr
\vskip.3cm
\noindent
We give the quantum analogue of a recently introduced electron model
which generalizes the Hubbard model with additional correlated hopping
terms and electron pair hopping. The model contains two independent
parameters and is invariant with respect to the quantum superalgebra
$U_q(gl(2|1))$. It  is integrable in one
dimension by means of the quantum inverse scattering method.

\+\cr
\+\cr
\vskip.5cm
\subheading{PACS numbers:} 71.20.Ad, 75.10.Jm

\vfil
\eject
\+\cr

In a recent letter \cite {1}, a new model of correlated electrons was
proposed which is integrable in one dimension on a periodic lattice by
means of the quantum inverse scattering method. The local Hamiltonian
was derived from a {\it rational} $R$-matrix satisfying the Yang-Baxter
equation and the system is invariant with respect to the Lie
superalgebra $gl(2|1)$. The resulting model generalizes the Hubbard
model with additional correlated hopping terms and electron pair
hopping and contains one free parameter. Solution of this model
with periodic boundary conditions has been investigated in \cite 2 by
application of the co-ordinate Bethe ansatz. {\it Quantization} of this
model, to obtain a system derived
 from a {\it trigonometric} $R$-matrix which has quantum
superalgebra $U_q(gl(2|1))$ invariance, introduces a further parameter
$q$ (the deformation or quantization parameter); this $R$-matrix has
already been given in general terms in \cite {1,3}. In this way we obtain
from \cite 1 a model with two free parameters and $U_q(gl(2|1))$
symmetry.

There is in fact an infinite number of abstract integrable models which
depend upon two generic  complex parameters. These are models associated
with the type I quantum superalgebras, which  admit
non-trivial one parameter families of representations and in turn  provide
solutions to the Yang-Baxter equation with additional parameteters
\cite {3-5}. From these solutions one may construct an integrable model
depending upon two (and possibly more) independent parameters.
As an example, we will construct in this Letter the quantized
version of the model \cite 1.

A two parameter generalization of the model \cite 1 has already
been proposed in
\cite {6}. Here we will show that the Hamiltonian of \cite 6
is  the quantized
analogue of that in \cite 1.
The Hamiltonian in \cite 6
reads
$$\align
H=&-\sum_{j,\si}(c_{j\si}^{\dag}c_{j+1\si} +h.c.)\exp\left [
-\frac 12(\eta -\si\ga)n_{j,-\si}-\frac 12(\eta+\si\ga)n_{j+1,-\si}
\right ] \\
&+\sum_j\left [Un_{j\upa}n_{j\doa}+t(c^{\dag}_{j\upa}c^{\dag}_{j\doa}
c_{j+1\doa}c_{j+1\upa}+h.c.) \right ],  \tag   {1}
\endalign   $$
where $j$ denotes the sites and the standard notation for the fermion
operators is used.
Integrability of the model with periodic boundary conditions
was established in \cite {6} under the
constraint
$$t=\frac {U}{2}=\ve\left [2e^{-\eta}(cosh\, \eta-cosh\,\ga)\right ]^{
\frac 12},\quad \ve =\pm 1 \tag {2}$$
leaving two free parameters $\eta,\,\ga.$ The imposition of periodic
boundary conditions breaks the quantum supersymmetry of the model.
However, we will show that the Hamiltonian with an additional chemical
potential term becomes $U_q(gl(2|1))$ invariant on the open lattice and
so is a particular case of the exactly solvable models with type I
quantum supersymmetry to which we referred above.

For the choice  $\ga=0$, the
above model is equivalent to the model \cite 1.
In order to derive the above  from \cite 1 through the quantum inverse
scattering method, we use the trigonometric $R$-matrix which is
invariant with respect to the one parameter family of four dimensional
representations of $U_q(gl(2|1))$.
Let
$\{\f|x\>\}_{x=1}^4$ denote an orthonormal
basis for a 4-dimensional $U_q(gl(2|1))$
module $V$. The quantum superalgebra $U_q(gl(2|1))$  has generators
$\{E^i_j\}_{i,j=1}^3$ which act on this module according to
$$\align
E^1_2&=\f|2\>\<3\r|,\quad E^2_1=\f|3\>\<2\r|,\quad E^1_1=-\f|3\>\<3\r|
-\f|4\>\<4\r|,\quad
E^2_2=-\f|2\>\<2\r|-\f|4\>\<4\r|,   \\
E^2_3&=[\al]_q^{\frac 12}\f|1\>\<2\r|+[\al+1]_q^{\frac 12}\f|3\>\<4\r|,\quad
E^3_2=[\al]_q^{\frac 12}\f|2\>\<1\r|+[\al +1]_q^{\frac 12}\f|4\>\<3\r|,  \\
E^1_3&=-[\al]_q^{\frac 12}\f|1\>\<3\r|+[\al+1]_q^{\frac 12}\f|2\>\<4\r|,\quad
E_1^3=-[\al]_q^{\frac 12}\f|3\>\<1\r|+[\al+1]_q^{\frac 12}\f|4\>\<2\r|, \\
E_3^3&=\al\f|1\>\<1\r|+(\al+1)(\f|2\>\<2\r|+\f|3\>\<3\r|)+
(\al+2)\f|4\>\<4\r|,   \tag {3}
\endalign     $$
where
$$[x]_q=\frac {q^x-q^{-x}}{q-q^{-1}},\qquad x\in\Bbb C. $$
Note that the representation depends upon a free parameter $\al\in\Bbb
C.$

The quantum superalgebra $U_q(gl(2|1))$ also carries a $\Z_2$-grading
determined by
$$\left [E^i_j\right ]=([i]+[j])(mod\,\,2)   \tag {4}  $$
where $[1]=[2]=0,\,[3]=1$. For $\al>0$ we have
$$\left (E^i_j\right )^{\dag}=E_i^j    $$
and we call the representation unitary of type I. For $\al<-1$ we have
$$\left (E_j^i\right )^{\dag}=(-1)^{[i]+[j]}E_i^j    $$
and we say the representation is unitary of type II. Hereafter we assume
that $\al $ is restricted to either of the above ranges. For a discussion
and classification of the unitary representations see \cite {7}.

Associated with $U_q(gl(2|1))$ there is  a co-product structure
($\Z_2$-graded algebra homomorphism) $\Del:U_q(gl(2|1))\rightarrow
U_q(gl(2|1))\ot U_q(gl(2|1))$ given by
$$\align
\Del(E^i_i)&=I\ot E_i^i +E_i^i\ot I,    \\
\Del(E_j^i)&=E_j^i\ot q^{\frac 12(E^i_i-E^j_j)}+q^{-\frac
12(E_i^i-E_j^j)}\ot E_j^i,   \qquad i<j,   \\
\Del(E_j^i)&=E_j^i\ot q^{\frac 12(E_j^j-E_i^i)}+ q^{-\frac 12
(E_j^j-E_i^i)} \ot E_j^i,\qquad i>j. \tag {5}
\endalign   $$
Under the co-product action the tensor product $V\ot V$ is also a
$U_q(gl(2|1))$ module which reduces completely;
$$V\ot V=V_1\oplus V_2\oplus V_3,   $$
where $V_1$ and $V_3$ are 4-dimensional modules and $V_2$ is
8-dimensional. Let $P_k,\,\,k=1,2,3$ denote projection operators from
$V\ot V$ onto $V_k$. There is a trigonometric $R$-matrix $\ch R(u)$
acting on $V\ot V$ which satisfies the Yang-Baxter equation
$$(I\ot \ch R(u))(\ch R(u+v)\ot I)(I\ot \ch R(v))=(\ch R(v)\ot I)(I \ot
\ch R(u+v))(\ch R(u)\ot I),     $$
and has the form  \cite {1}
$$\ch R(u)=\frac {q^u-q^{2\al}}{1-q^{u+2\al}}P_1+P_2+\frac{1-q^{u+2\al
+2}}{q^u-q^{2\al+2}}P_3.   $$
We point out that multiplication of tensor products is governed
by
$$(a\ot b)(c\ot d)=(-1)^{[b][c]}(ac\ot bd).    $$

The projection operators may be evaluated as follows. Let $\f|\Psi^1_k\>,
\,\f|\Psi^3_k\>,\,k=1,2,3,4$ form symmetry adapted bases for the spaces
$V_1$ and $V_3$ respectively. By means of the representation (3) and the
co-product action (5) one may verify that
$$\align
\f|\Psi^1_1\>=&\f|1\>\ot \f|1\>,    \\
\f|\Psi^1_2\>=&(q^{\al}+q^{-\al})^{-\frac 12}(q^{\frac 12\al}\f|1\>\ot\f|2\>+
q^{-\frac 12\al}\f|2\>\ot\f|1\>), \\
\f|\Psi^1_3\>=&(q^{\al}+q^{-\al})^{-\frac 12}(q^{\frac 12\al}\f|1\>\ot\f|3\>+
q^{-\frac 12\al}\f|3\>\ot \f|1\>,   \\
\f|\Psi^1_4\>=&(q^{\al}+q^{-\al})^{-\frac 12}[2\al+1]_q^{-\frac 12}\left [
[\al+1]_q^{\frac 12}(q^{\al}\f|1\>\ot\f|4\>+q^{-\al}\f|4\>\ot\f|1\>)\right. \\
&\qquad\qquad \left. +[\al]_q^{\frac
12}(q^{\frac 12}\f|2\>\ot \f|3\>-q^{-\frac 12}\f|3\>\ot \f|2\>)\right ], \\
\f|\Psi^3_1\>=&(q^{\al+1}+q^{-\al-1})^{-\frac 12}[2\al+1]_q^{-\frac 12} \left
[[\al]_q^{\frac 12}(q^{\al+1}\f|4\>\ot\f|1\>+q^{-\al-1}\f|1\>\ot\f|4\>)
\right.  \\
&\qquad\qquad\left. +[\al+1]_q^{\frac 12}(q^{-\frac 12}\f|3\>\ot\f|2\>
-q^{\frac 12}\f|2\>\ot\f|3\>) \right ],    \\
\f|\Psi^3_2\>=&(q^{\al+1}+q^{-\al-1})^{-\frac 12}(q^{\frac 12(\al+1)}
\f|4\>\ot\f|2\>+ q^{-\frac 12(\al+1)}\f|2\>\ot\f|4\>),   \\
\f|\Psi^3_3\>=&(q^{\al+1}+q^{-\al-1})^{-\frac 12}(q^{\frac
12(\al+1)}\f|4\>\ot\f|3\>+q^{-\frac 12(\al+1)}\f|3\>\ot\f|4\>),   \\
\f|\Psi^3_4\>=&\f|4\>\ot\f|4\>
\endalign   $$
and this basis is orthonormal. Consistent with the $\Z_2$-grading (4) we
may assign a grading on the basis states  by
$$[\f|1\>]=[\f|4\>]=0,\qquad[\f|2\>]=[\f|3\>]=1.   \tag {6} $$
Using the rules
$$\left (\f|x\>\ot\f|y\>\right)^{\dag}=(-1)^{[|x>][|y>]}\<x\r|\ot\<y\r|,$$
$$\<\Psi_k^i\r|=\f|\Psi^i_k\>^{\dag},\qquad k=1,2,3,4,\quad i=1,3$$
we have
$$\align
P_1&=\sum_{k=1}^4\f|\Psi^1_k\>\<\Psi_k^1\r|,\qquad
P_3=\sum_{k=1}^4\f|\Psi_k^3\>\<\Psi_k^3\r|,  \\
P_2&=I-P_1-P_3.     \endalign   $$

On the $N$-fold tensor product space we denote
$$\ch R(u)_{i,i+1}=I^{\ot (i-1)}\ot\ch R(u)\ot I^{(N-i-1)},   $$
and define a local Hamiltonian by \cite 8
$$\align
H_{i,i+1}&=-\frac{(q^{\al+1}-q^{-\al-1})}{ln\,q}\left.\frac {d}{du} \ch
R(u)_{i,i+1}\right |_{u=0}  \\
&=(q^{\al}+q^{-\al})\frac{[\al+1]_q}{[\al]_q}(P_1)_{i,i+1}-
(q^{\al+1}+q^{-\al-1} )(P_3)_{i,i+1}.
\endalign  $$
In view of the grading (6) we now make the assignment
$$\align
&\f|4\>\equiv \f|0\>,\qquad\qquad\qquad \f|3\>\equiv
\f|\upa\>=c^{\dag}_{\upa}\f|0\>,\\
&\f|2\>\equiv \f|\doa\>=c^{\dag}_{\doa}\f|0\>, \qquad \f|1\>\equiv
\f|\upa\doa\>=c_{\doa}^{\dag}c_{\upa}^{\dag}\f|0\>,  \endalign
$$
which allows us to express the projection operators in terms of the
canonical fermion operators. These expressions are complicated, so we
will just present the local Hamiltonian which reads
$$\align
H_{i,i+1}=&\sum_{\si}(c^{\dag}_{i\si}c_{i+1\si} +h.c.)\left[ -1+n_{
i,-\si}(1+\nu  q^{\frac 12\si}\left(\frac {[\al+1]_q}{[\al]_q}
\right )^{\frac 12})\right.     \\
&\qquad +n_{i+1,-\si}(1+\nu q^{-\frac
12\si}\left(\frac{[\al+1]_q}{[\al]_q}\right)^{\frac 12})  \\
& \qquad    \left. -n_{i,-\si}n_{i+1,-\si}(1+\frac{[\al+1]_q}{[\al]_q}
 +\nu (q^{\frac 12} +q^{-\frac 12})\left(\frac{
[\al+1]_q}{[\al]_q}\right)^{\frac 12}) \right ]   \\
&  \qquad     +[\al]_q^{-1}(c_{i_\doa}^{\dag}c_{i\upa}^{\dag}c_{i+1\upa}
c_{i+1\doa} +h.c.)    \\
&\qquad +[\al]_q^{-1}(n_{i\upa}n_{i\doa} +n_{i+1\upa}n_{i+1\doa})-(q^{\al+1}
+q^{-\al-1})      \\
&\qquad +q^{\al+1}(n_{i\upa}+n_{i\doa})+q^{-\al-1}(n_{i+1\upa}+n_{i+1\doa}),
\endalign    $$
where $\nu=sgn\,\al$. Under the unitary transformation
$$c_{i\si}\rightarrow c_{i\si}(1-2n_{i,-\si})$$
we yield the same local Hamiltonian with $\nu=-sgn\,\al$.
Observe also that the local Hamiltonian is hermitian only for $\al>0$ or
$\al<-1$; i.e. when the underlying representation is unitary.

On the periodic
lattice we have that $H=\sum_iH_{i,i+1}$ is equivalent to the Hamiltonian (1)
under the constraint (2) (up to a constant term and chemical potential
term) by choosing
$$e^{\ga}=q,\qquad e^{-\eta}=\frac{[\al+1]_q}{[\al]_q}$$
which gives $U=2[\al]^{-1}$. Ref. \cite 6 establishes integrability of
the model on
the periodic chain although for such a case there is no quantum
superalgebra symmetry. From the results of \cite {9} we can also deduce
integrability on the open chain which gives a model with $U_q(gl(2|1))$
invariance.     Detailed results will be deferred to
a separate publication.

In conclusion, we have indicated in this Letter that integrable  models
dependent upon two generic parameters may be naturally constructed
through the quantum inverse scattering method and $R$-matrices
associated with the type I quantum superalgebras. As an example we have
considered one of the simplest cases and derived  the quantum analogue
of an electron model with
correlated hopping \cite 1.
Using the $R$-matrix, we have obtained the explicit
form of the Hamiltonian. This approach can  be extended to
generate higher conservation laws (c.f.
\cite {10} for the case of the supersymmetric
$t-j$ model).
\+\cr
\+\cr
\subheading{Acknowledgement} We are deeply indebted to A.J. Bracken
for helpful suggestions on improving the first draft of the manuscript.
Financial support from the Australian Research Council is
greatly appreciated. YZZ is supported by the Kyoto University
Foundation.

\vfil
\eject

\heading References \endheading
\vs
\ref\no 1\by A.J. Bracken, M.D. Gould, J.R. Links and Y.-Z. Zhang
\jour Phys. Rev. Lett. \vol 74 \yr 1995  \page 2768  \endref

\ref\no 2\by G. Bed\"urftig and H. Frahm \paper Thermodynamics of an
integrable model for electrons with correlated hopping \jour
cond-mat/9504103  \endref

\ref\no 3\by A.J. Bracken, G.W. Delius, M.D. Gould and Y.-Z. Zhang
\jour J. Phys. A: Math. Gen. \vol 27 \yr 1994 \page 6551 \endref

\ref\no 4\by G.W. Delius, M.D. Gould, J.R. Links and Y.-Z. Zhang \paper
On type I quantum affine superalgebras \jour hep-th/9408006, Int. J.
Mod. Phys. \vol A \toappear   \endref

\ref\no 5\by G.W. Delius, M.D. Gould, J.R. Links and Y.-Z. Zhang \paper
 Solutions of the Yang-Baxter equation with extra non-additive
 parameters II: $U_q(gl(m|n))$ \jour hep-th/9411241 \endref

\ref\no 6\by R.Z. Bariev, A. Kl\"umper and J. Zittartz \paper A new
integrable two parameter model of strongly correlated electrons in one
dimension \jour cond-mat/9504114    \endref

\ref\no 7\by M.D. Gould and M. Scheunert \jour J. Math. Phys. \yr 1995
\vol 36 \page 435    \endref

\ref\no 8\by P. Kulish and E. Sklyanin \jour Lect. Notes in Phys. \vol
151 \yr 1982 \page 61 \endref

\ref\no 9\by J.R. Links and M.D. Gould \paper Integrable open chains
with quantum supersymmetry \jour in preparation  \endref

\ref\no 10\by F.H.L. Essler and V.E. Korepin \jour Phys. Rev. \vol B46
\yr 1992 \page 9147   \endref

\enddocument